\documentclass[reprint,superscriptaddress,groupedaddress,aps,prl,twocolumn]{revtex4-2}
\usepackage[T1]{fontenc}
\usepackage[latin1]{inputenc}
\usepackage{lmodern}
\expandafter\let\csname equation*\endcsname\relax
\expandafter\let\csname endequation*\endcsname\relax
\usepackage{amsmath}
\usepackage{amssymb}
\usepackage{float}
\usepackage{graphicx}
\usepackage{xcolor}
\usepackage{natbib}
\usepackage{bm}
\usepackage{dsfont}
\usepackage{soul}
\usepackage{bbold}
\usepackage[mathscr]{euscript}
\usepackage[cal=boondoxo]{mathalfa}
\usepackage{comment}
\usepackage{notes2bib}
\bibnotesetup{
note-name = ,
use-sort-key = false
}
\renewcommand\footnotemark{}

\usepackage{braket}
\usepackage[normalem]{ulem}
\def\ii{{\rm i}}

\def\hge{\hat{\sigma}_{ge}}  
\def\heg{\hat{\sigma}_{eg}}

\def\ga{\Gamma_{\rm 1D}}
\def\kk{k_{\rm 1D}}
\def\gap{\Gamma'}

\def\ra{\hat{\rho}}

\def\bra#1{\mathinner{\langle{#1}|}}
\def\ket#1{\mathinner{|{#1}\rangle}}
\def\braket#1{\mathinner{\langle{#1}\rangle}}

\def\gp{\Gamma_+} 
\def\gm{\Gamma_-} 

\def\opn{\hat{\mathcal{O}}_\nu}

\def\gl{\Gamma_L}
\def\gr{\Gamma_R} 
\def\jop{\hat{\mathcal{O}}}
\def\g2{g^{(2)}}

\begin{document}
\title{Many-body superradiance and dynamical mirror symmetry breaking in waveguide QED}
\author{Silvia Cardenas-Lopez, Stuart J. Masson, Zoe Zager, and Ana Asenjo-Garcia}
\email{ana.asenjo@columbia.edu}
\affiliation{Department of Physics, Columbia University, New York, NY 10027, USA}
\date{\today}
\begin{abstract}
The many-body decay of extended collections of two-level systems remains an open problem. Here, we investigate whether an array of emitters coupled to a one-dimensional bath undergoes Dicke superradiance, a process whereby a completely inverted system becomes correlated via dissipation. This leads to the release of all the energy in the form of a rapid photon burst. We derive the minimal conditions for the burst to happen as a function of the number of emitters, the chirality of the waveguide, and the single-emitter optical depth, both for ordered and disordered ensembles. Many-body superradiance occurs because the initial fluctuation that triggers the emission is amplified throughout the decay process. In one-dimensional baths, this avalanche-like behavior leads to a spontaneous mirror symmetry breaking, with large shot-to-shot fluctuations in the number of photons emitted to the left and right. Superradiant bursts may thus be a smoking gun for the generation of correlated photon states of exotic quantum statistics. 
\end{abstract}
\maketitle
The decay rate of a single emitter is dictated by its radiative environment~\cite{Purcell46,Kleppner81,Haroche89}. This realization contributed to the development of cavity quantum electrodynamics (QED). Here, highly-reflecting mirrors isolate a single optical mode, yielding a localized (or zero-dimensional) reservoir for the emitter, which enhances its decay into the cavity. One-dimensional (1D) baths pertain to ``waveguide QED'', where an atom is interfaced with a propagating optical mode. Recent years have seen tremendous experimental progress, with platforms including cold atoms coupled to optical nanofibers~\cite{Vetsch10, Goban12, Gouraud15, Solano17,Liedl22}, cold atoms~\cite{Thompson13,Goban15,Hood16} or quantum dots~\cite{Lodahl15,Tiranov23} coupled to photonic crystal waveguides, and superconducting qubits coupled to microwave transmission lines~\cite{Liu16,Mirhosseini19,Zanner2022}. Besides altering decay, interfacing several emitters with 1D propagating modes allows engineering of long-range atom-atom interactions~\cite{LeKien08,chiral,PhysRevResearch.3.033233,Albrecht_2019}.

The environment also determines the many-body decay of a multiply-excited ensemble. A paradigmatic example of many-body decay is Dicke superradiance: a collection of fully-inverted emitters phase-locks as they decay, emitting a short bright pulse of photons~\cite{Dicke54,Gross82}. Despite its many-body nature, this problem is solvable in a cavity due to the imposed permutational symmetry restricting dynamics to a small subset of states of the (otherwise exponentially-large) Hilbert space. In extended systems, atom-atom interactions depend on their positions and many-body decay generates complex dynamics ~\cite{Friedberg72,Skribanowitz73,Flusberg76,Gross76,Rubies21,Robicheaux21,Ferioli21,Masson22,Sierra22,Orell22,Wiegner11} that remains to be fully understood.

In free space, superradiance can generate highly directional emission due to the sample geometry~\cite{Rehler71,Gross82,Clemens2003,Inouye99}. Symmetry in emission direction can also be broken spontaneously, as atom-atom correlations lead to ``memory effects'': detecting a photon in one specific direction increases the likelihood for that detector to record subsequent photons. As the direction of emission of the first photon is random, and due to the avalanche-like nature of the process, Dicke superradiance has been predicted to give rise to large shot-to-shot fluctuations in the angular distribution of the far field intensity, both for atomic ensembles~\cite{Dicke54,Carmichael2000,Clemens2004,Liberal19} and Bose-Einstein condensates \cite{Moore99}. However, the many optical field modes make the theoretical analysis of this phenomenon challenging.

Here, we investigate the decay of a fully inverted array of emitters into a 1D bath. Leveraging previous work~\cite{Masson22} to bypass the exponential growth of the Hilbert space by studying early dynamics, we set constraints on the number of emitters needed to observe superradiance. We investigate chiral and bidirectional waveguides interfaced with ordered and disordered ensembles. In 1D, due to the confined nature of the optical fields, the build-up of directional correlations translates into spontaneous breaking of mirror symmetry, giving rise to an emergent chirality. We compute the probability distribution of directional emission, which is very broad due to large shot-to-shot fluctuations. The distribution is shown to evolve with time, as Hamiltonian evolution scrambles the correlations imprinted by the dissipative process, thus washing away the memory of the system. This physics can be explored in diverse experimental setups.

\begin{figure}
\centering
\includegraphics[width=0.9\linewidth]{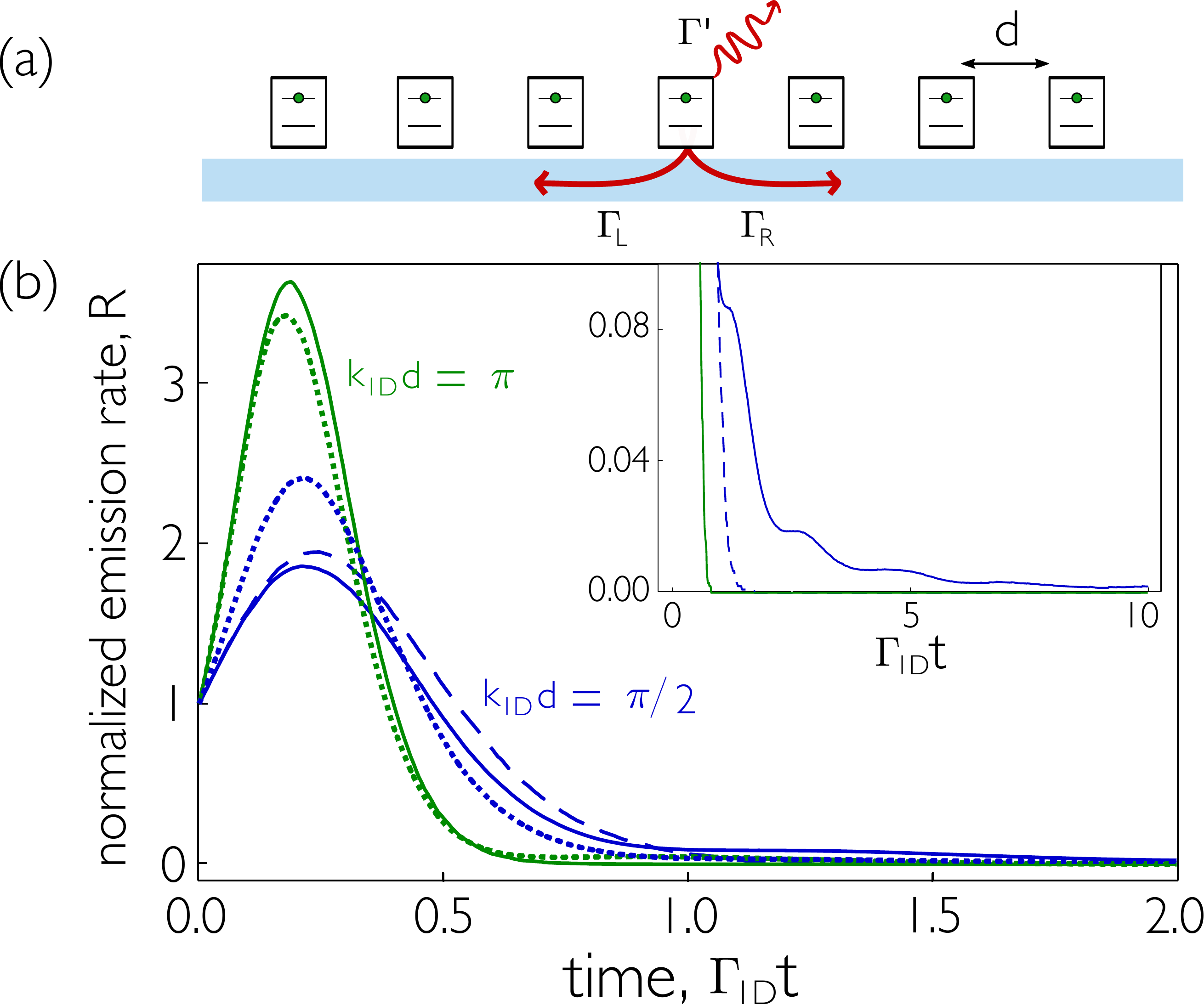}
\caption{Many-body superradiance from emitters coupled to a waveguide. (a) Schematic: $N$ emitters of lattice constant $d$ interact via a 1D bath, which supports propagation of photons of wavevector $\pm\kk$. Single-emitter decay rates into left and right-propagating modes of the waveguide are $\Gamma_{L/R}$ respectively, and any other parasitic decay is denoted by $\Gamma'$. (b) Emission rate into the waveguide for an array of $N=16$ emitters coupled to a bidirectional (solid lines) and a chiral (dotted) waveguide with $\gr=3\gl$ and $\Gamma'=0$. Dashed line shows the bidirectional waveguide calculation without Hamiltonian contribution, which is significant at late times (inset).}
\label{fig:fig1}
\end{figure} 

We consider $N$ emitters of resonance frequency $\omega_0$ coupled to a 1D photonic channel, as shown in Fig.~\ref{fig:fig1}(a). The waveguide mode mediates interactions between emitters. Tracing out the photonic degrees of freedom under a Born-Markov approximation, the evolution of the emitters' density matrix in the rotating frame is described by the master equation~\cite{Gruner96,Dung02}
\begin{equation}
\dot{\ra}=-\frac{\ii}{\hbar}\left[\mathcal{H}_L+\mathcal{H}_R,\ra\right]+\mathcal{L}_g[\ra]+\mathcal{L}_{ng}[\ra].
\label{eq:master}
\end{equation}
Here, the Hamiltonians $\mathcal{H}_{L/R}$ allow for distinct coupling to left- and right-propagating waveguide modes (at rates $\Gamma_{L/R}$ for a single emitter), and read~\cite{Pichler15}
\begin{subequations}
\begin{gather}
\mathcal{H}_L=-\frac{\ii\hbar \Gamma_L}{2}\sum_{i<j}e^{\ii \kk |z_i-z_j|}\heg^i\hge^j+\text{H.c.},\\
\mathcal{H}_R=-\frac{\ii\hbar \Gamma_R}{2}\sum_{i>j}e^{\ii \kk |z_i-z_j|}\heg^i\hge^j+\text{H.c.},
\end{gather}\label{ham}
\end{subequations}
where $\hge^i=\ket{g_i}\bra{e_i}$ is the coherence operator between the ground and excited states of emitter $i$ at position $z_i$, $\kk$ is the photon wavevector, and H.c. stands for Hermitian conjugate. The total decay rate of a single emitter into the waveguide is $\ga=\Gamma_L+\Gamma_R$. The Lindblad operators $\mathcal{L}_g[\ra]$ and $\mathcal{L}_{ng}[\ra]$ describe the decay of emitters to guided and non-guided modes, respectively, and read
\begin{equation}
\mathcal{L}_\alpha[\ra]=\sum_{i,j=1}^N\frac{\Gamma^\alpha_{ij}}{2}\left(2\hge^j\ra\heg^i-\ra\heg^i\hge^j-\heg^i\hge^j\ra\right),
\label{eq:master2}
\end{equation}
where $\Gamma^{g}_{ij}=\Gamma_L e^{\ii \kk (z_j-z_i)}+\Gamma_R e^{-\ii \kk (z_j-z_i)}$ and $\Gamma^{ng}_{ij}=\Gamma'\delta_{ij}$. We consider that non-guided decay is not collective, either because it represents local parasitic decay or because emitters are far separated and interactions via non-guided modes are negligible.

Emission of photons into the waveguide is correlated due to the shared bath. This is captured by collective jump operators found by diagonalizing the $N\times N$ Hermitian matrix $\mathbb{\Gamma}$ of elements $\Gamma_{ij}^g$ \cite{Carmichael2000,Clemens2003}.
Photons can only be emitted into the left- or right-propagating modes, and thus $\mathbb{\Gamma}$ has only two non-zero eigenvalues and we can write
\begin{equation}
\mathcal{L}_g[\ra]=\sum_{\nu=+,-}\frac{\Gamma_\nu}{2}\left(2\opn\ra\,\opn^\dagger-\ra\,\opn^\dagger\opn-\opn^\dagger\opn\ra\right),
\label{eq:guided}
\end{equation}
where $\hat{\mathcal{O}}_\nu$ are collective jump operators and $\Gamma_\nu$ are collective decay rates, found as the eigenvectors and eigenvalues of $\mathbb{\Gamma}$ respectively. The $\{+,-\}$ notation indicates that $\hat{\mathcal{O}}_{+(-)}$ generates a photon in a symmetric (antisymmetric) superposition of left- and right-propagating modes.

A fully inverted initial state, $\ket{\psi(t=0)}=\ket{e}^{\otimes N}$, will decay due to vacuum fluctuations, leading to emission into the waveguide at a (normalized) rate 
\begin{equation}
    R(t)=\frac{1}{N \ga}\sum_{\nu=+,- }\Gamma_\nu \braket{\hat{\mathcal{O}}_\nu^\dagger\hat{\mathcal{O}}_\nu}.
\end{equation}
For large enough $N$ and $\gap=0$, a superradiant burst occurs for any lattice constant, as shown in Fig.~\ref{fig:fig1}(b) for ordered arrays. Calculations are performed using quantum trajectories~\cite{Dalibard92,Carmichael93,SI}. Maximal superradiance occurs in a bidirectional waveguide (i.e., $\Gamma_R=\Gamma_L)$, at the so-called ``mirror configuration'' ($\kk d=n\pi$ with $n\in \mathbb{N}$~\cite{Chang12,Corzo16,Sorensen16}), as this situation corresponds to that studied by Dicke.

Dissipative dynamics are the main driving mechanism for the burst. The coherent (i.e., Hamiltonian) interactions only contribute well beyond the time of maximum emission. At later times the Hamiltonian plays a significant role, leading to oscillations in emission as it cycles the atoms between dark and bright states, [see inset to Fig.~\ref{fig:fig1}(b)].
\begin{figure*}
\centering
\includegraphics[width=1\textwidth]{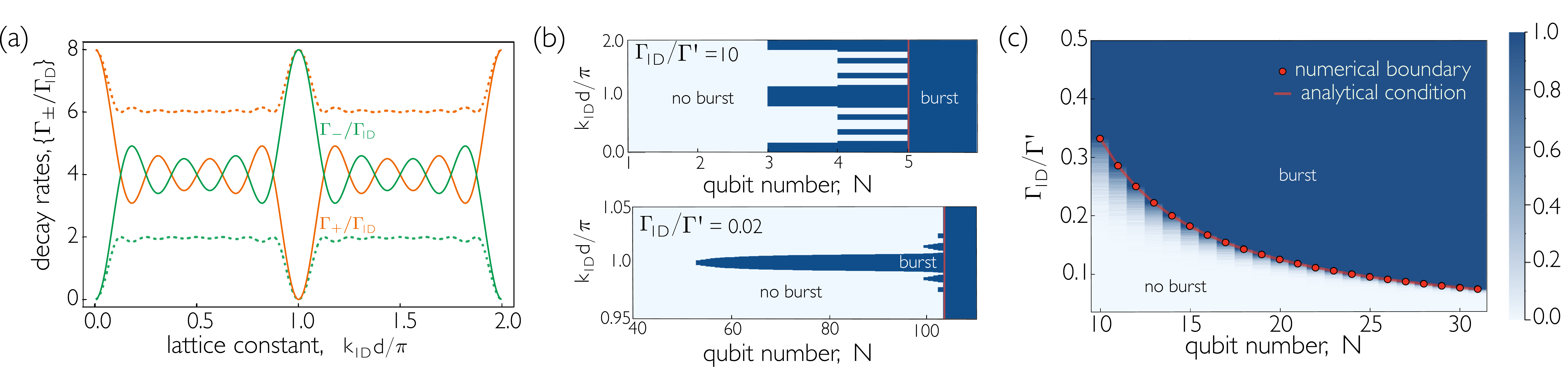}
\caption{Predictions of superradiance for ordered and disordered ensembles. (a) Collective decay rates into a bidirectional (solid lines) and a chiral (dotted) waveguide for $N=8$. Single-emitter decay rates for the chiral waveguide are $\Gamma_R = 3\Gamma_L$.
(b)~Crossover between burst (dark blue) and no-burst (pale blue) regions in ordered arrays coupled to a bidirectional waveguide. (c)~Probability of having a burst in a bidirectional waveguide for spatially-disordered ensembles of emitters randomly placed over a section of length $\kk z_\text{max}\gg 2\pi$~\bibnotemark[SI]. In (b,c) the red line shows the emitter number that guarantees a burst regardless of separation, as given by Eq.~\eqref{fin}.}
\label{fig:fig2}
\end{figure*}

As we postulated in prior work~\cite{Masson22}, the minimal condition for a burst is that the first photon enhances the emission of the second. The same insight can be used to derive a condition for a superradiant burst into one particular channel, even if emitters can decay to more than one reservoir (as happens in the presence of finite $\Gamma'$). To do so, we adapt the calculation from Ref.~\cite{Robicheaux21} of ``directional superradiance'' and define a second-order correlation function conditioned on measurement only of the waveguide modes
\begin{equation}
    \tilde{g}^{(2)}(0) = \frac{\sum\limits_{\nu=\pm} \sum\limits_{\mu=\pm,i} \Gamma_\nu \Gamma_\mu \braket{\jop^\dagger_\mu \jop^\dagger_\nu \jop_\nu \jop_\mu}}{\left(\sum\limits_{\nu=\pm}\Gamma_\nu \braket{\jop^\dagger_\nu\jop_\nu}\right) \left(\sum\limits_{\mu=\pm,i} \Gamma_\mu \braket{\jop^\dagger_\mu\jop_\mu}\right)}.
    \label{eq:g2}
\end{equation}
Here, sums over $\pm$ account for waveguide emission, while sums in $i$ account for local decay. The minimal condition for superradiant emission of photons into the waveguide is found by imposing $\tilde{g}^{(2)}(0)>1$. Importantly, this condition selects the processes in which the photon emission rate is enhanced into the waveguide only.

For an ensemble of initially-inverted emitters this condition becomes [see Supplemental Material (SM)~\cite{SI}] 
\begin{equation}
\mathrm{Var}\left(\frac{\{\Gamma_\nu\}}{\ga}\right) > 1+\frac{\gap}{\ga},
\label{cond}
\end{equation}
where $\mathrm{Var}(\cdot)$ is the variance~\bibnote[SI]{See Supplemental Material for further details on the second-order correlation function, the emergent translational symmetry at the degeneracy points, chirality induced by jumps, imbalance statistics in the presence of parasitic decay, superradiance in giant atoms and details on the numerical evolution.}. This expression is general; it applies to systems with any number of emitters, in disordered or ordered spatial configurations, and coupled to waveguides with any degree of chirality. A burst occurs if there are only a few dominant decay channels (maximizing the variance), and if collective decay overcomes local loss. As emission is constrained to 1D, there are at most two bright channels, while $N-2$ are dark (i.e., of zero decay rate). Therefore, the conditions for a burst are more easily satisfied than for arrays in free space~\cite{Masson20,Masson22,Sierra22}. Restriction of emission into a 1D bath eliminates most of the competition between different imprinted phase patterns, enabling a more robust phase-locking than in free space, where photons can be emitted in all directions.

For ordered arrays of lattice constant $d$, the two collective decay rates admit the analytical form 
\begin{equation}
\Gamma_\pm=\frac{N\ga}{2}\pm \sqrt{\frac{N^2(\Gamma_L-\Gamma_R)^2}{4}+\Gamma_L\Gamma_R\frac{\sin^2 N\kk d}{\sin^2 \kk d}}.
\label{gammas_nu}
\end{equation}
The two decay rates are generally distinct and finite, as shown in Fig. \ref{fig:fig2}(a), leading to competition between the $\pm$ channels. Different lattice constants give rise to situations ranging from the Dicke model with a single non-zero collective decay rate to maximum competition, where the decay rates are degenerate due to an emergent translational symmetry~\cite{SI}. For chiral waveguides there is no degeneracy, as any level of chirality breaks translation symmetry. In this case, rates are almost independent of the lattice constant, as interference effects are suppressed.

The minimal burst condition for ordered arrays reads
\begin{equation}
    \frac{N\left(\Gamma_L^2+\Gamma_R^2\right)}{\ga^2}+2 \frac{\Gamma_L\Gamma_R}{N\ga^2}\frac{\sin^2 N\kk d}{\sin^2 \kk d}> 2+\frac{\gap}{\ga}.
    \label{finalcond}
\end{equation}
Large parasitic decay quenches the superradiant burst for small $N$, as shown in Fig. \ref{fig:fig2}(b). However (and regardless of the level of independent decay) a burst is always recovered if the number of emitters is increased beyond a certain threshold.

For disordered systems we obtain the minimal burst condition in terms of single-emitter decay rates by placing a lower bound on the trace of $\mathbb{\Gamma}^2$, as the eigenvalues do not admit an analytical form. As demonstrated in~\bibnotemark[SI], $\text{Tr}[\mathbb{\Gamma}^2]\geq  N^2 (\Gamma_R^2+\Gamma_L^2)$, and the burst is guaranteed to happen for 
\begin{align}
\frac{N(\Gamma_L^2+\Gamma_R^2)}{\ga^2}>2+\frac{\gap}{\ga}.
    \label{fin}
\end{align}
Disordered systems saturate this bound [see Fig~\ref{fig:fig2}(c)], while ordered systems may display a burst for lower $N$ due to interference effects [see Eq.~\eqref{finalcond} and Fig~\ref{fig:fig2}(b)]. 

\begin{figure*}
\centering
\includegraphics[width=0.95\textwidth]{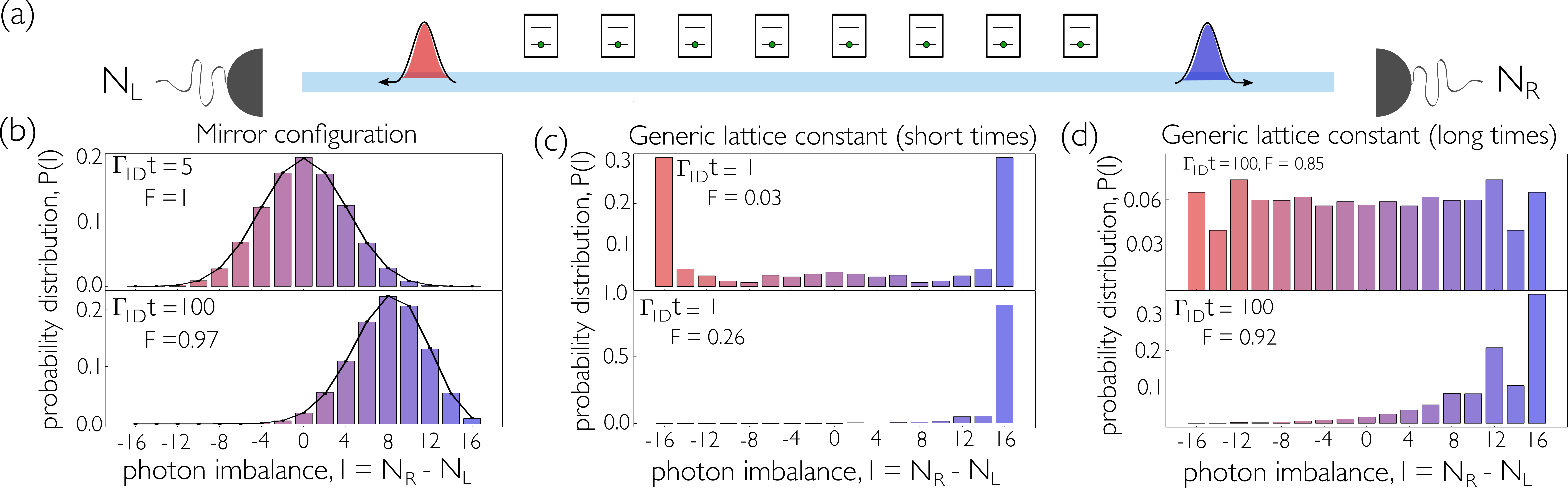}
\caption{Emergence of chirality via many-body decay. (a) Superradiant decay leads to directionality that can be investigated by detectors which count the number of right- and left-propagating photons. (b-d) Directional photon imbalance for 16 emitters radiating into a symmetric (top) and asymmetric (bottom) waveguide, with $\gap=0$. For the latter, $\gr=3\gl$. The lattice constant is (b) $\kk d=\pi$ and (c,d) $\kk d=\pi/\sqrt{3}$. In (c,d), we investigate short and long times respectively. Only trajectories that reach $\ket{g}^{\otimes N}$ before time $t$ are accounted for, with $F$ denoting the fraction of finished trajectories.}
\label{fig:fig3}
\end{figure*} 

Generically, correlations imparted by the jump operators not only produce an accelerated emission of the second photon, but also of subsequent ones. This avalanche-like nature of photon emission implies that an initial fluctuation is amplified throughout the decay process. Shot-to-shot fluctuations in directionality have been predicted in free space~\cite{Carmichael2000,Moore99}. In a 1D bath, however, these fluctuations are more striking as there are only two directions, and the fluctuations break mirror symmetry. For instance, if the first photon is measured by a detector to the right, it is very likely that subsequent photons are also detected to the right. This process gives rise to an emergent chirality even in the case of a bidirectional waveguide. To explore this physics, we unravel $\mathcal{L}_g[\rho]$ in terms of a different pair of operators,
\begin{equation}
    \hat{\mathcal{O}}_{L/R}=\frac{1}{\sqrt{N}}\sum_{i=1}^N e ^{\pm \ii \kk d i}\hge^i\propto \sqrt{\Gamma_+}\hat{\mathcal{O}}_{+}\pm\ii\sqrt{\Gamma_-}\hat{\mathcal{O}}_{-},
    \label{eq:left_right}
\end{equation}
which describe the emission of photons to left- and right-propagating modes with rates $\Gamma_{L/R}$.

The direction of the first photon is stochastic due to the uncorrelated initial state, with probability depending only on the relative decay rate of each operator, i.e., $p_{L/R}= \Gamma_{L/R}/\ga$. Emergent chirality is already evident at the level of two emissions. If all the photons are emitted into the waveguide (i.e., $\gap=0$), the detection probabilities for the second photon to be the same as or different to the first one are
\begin{subequations}
\begin{eqnarray}
     \tilde{g}_{LL}^{(2)}(0)&&=2-\frac{2}{N},\label{gll}\\
     \tilde{g}_{LR}^{(2)}(0)&&=1-\frac{2}{N}+\frac{1}{N^2}\frac{\sin^2 N \kk d}{\sin^2 \kk d}.\label{glr}
\end{eqnarray}
\end{subequations}
where $\tilde{g}_{\alpha\beta}^{(2)}(0)= \braket{\jop^\dagger_\alpha\jop^\dagger_\beta\jop_\beta\jop_\alpha}/(\braket{\jop^\dagger_\alpha\jop_\alpha}\braket{\jop^\dagger_\beta\jop_\beta})$.

For large $N$, the second photon is twice as likely to follow the direction of the first. Subsequent jumps further enhance the chirality~\cite{SI} (except in the mirror configuration where $\hat{\mathcal{O}}_{L/R}$ are identical). One can attribute this enhanced chirality to the correlations produced by atom-atom interactions or to the photon detection (far-field measurement is unable to distinguish which atom emitted the photon thus preparing a superposition state~\cite{Wiegner11,Liberal19,Bojer22}). 

For bidirectional waveguides, this emergent chirality is akin to a process of spontaneous symmetry breaking, where mirror symmetry is broken dynamically. A large superradiant burst implies that, for a single realization, most photons are emitted in one direction. Of course, the symmetry is recovered when averaged over realizations, as the first photon is randomly emitted into either direction.

We characterize this behavior by counting the photons emitted in both directions and computing the ``photon imbalance'', as shown in Fig.~\ref{fig:fig3}(a). For a single quantum trajectory -- in which atoms evolve from $\ket{e}^{\otimes N}$ to $\ket{g}^{\otimes N}$ via coherent evolution with a non-Hermitian Hamiltonian and decay by the action of jump operators -- we define the photon imbalance as the difference in the number of times that the two jump operators act. This corresponds to counting the final number of photons emitted to the right ($N_R$) and left ($N_L$) with imbalance $I=N_R-N_L$. The set of possible photon imbalances $I$ has a probability distribution, $\mathcal{P}(I)$. The imbalance distribution depends on the lattice constant and the degree of single-emitter chirality of the waveguide. 

In the mirror configuration, as the left and right operators are identical, the normalized probability of emitting a photon in each direction reduces to approximately $p_{L/R}$ at any stage of the decay. Hence, the photon imbalance roughly follows a binomial distribution, as shown in Fig.~\ref{fig:fig3}(b). 
Minor discrepancies originate from the action of the Hamiltonian in between jumps (only for a chiral waveguide) and noise from the finite number of trajectories. 

Away from the mirror configuration, the repeated action of a jump operator enhances its probability of acting again, thus amplifying the initial fluctuation and either breaking mirror symmetry (for a bidirectional waveguide) or collectively enhancing chirality (for a chiral one). Repeated action leads to emission that finishes at early times because it produces strongly enhanced photon emission, so the time between jumps is small and Hamiltonian evolution is negligible. This is shown in Fig.~\ref{fig:fig3}(c). For a bidirectional waveguide, almost all photons are emitted in one direction. A mildly-chiral waveguide becomes almost perfectly chiral.

Hamiltonian evolution becomes relevant for the imbalance statistics at later times \cite{Nha2002} by scrambling the states and reducing enhancement (yet still giving rise to a distribution with a very large variance). As shown in Fig.~\ref{fig:fig3}(d), for the bidirectional waveguide, it gives rise to an almost-flat imbalance distribution and for the chiral waveguide, the enhancement of the chirality is reduced. Nonetheless, the probability of detecting all photons in a single direction is much greater than the probability predicted by the binomial distribution for independent emission. This resembles a recent prediction for multilevel atoms in a cavity, where there is a higher probability of large imbalances between ground state populations compared to single-atom predictions~\cite{Pineiro22}.

In conclusion, we have established a condition for enhanced emission into a preferential channel when emitters decay to multiple reservoirs. We have found the minimal conditions for the emission of a superradiant burst into a 1D bath and determined that the burst should be observable in different experimental setups, such as superconducting qubits coupled to transmission lines and atoms coupled to nanofibers. Many-body superradiance gives rise to an emergent chirality in the system, with large amounts of photons being emitted in one direction. As shown in~\cite{SI}, large photon imbalances disappear with increased $\gap$. Nevertheless, pronounced imbalances should be observable in state-of-the-art experimental setups with superconducting qubits, where $\Gamma'\simeq 0.01\Gamma_\text{1D}$~\cite{Mirhosseini19}, and quantum dots, where $\Gamma'\simeq 0.1\Gamma_\text{1D}$~\cite{Tiranov23}.

Giant atoms (emitters coupled to the waveguide at multiple points~\cite{Kockum14,Kockum18,Kannan2020,Guo17,Longhi20,Du22}) also exhibit superradiance when the parasitic decay is smaller than the individual decay into the waveguide~\cite{SI}. The interference of each emitter with itself modifies the individual decay rate, and certain configurations result in a decoherence-free system with non-zero coherent interactions~\cite{Kockum18}. Atoms near these configurations can exhibit monotonic decay. Interestingly, the additional tunability of the coherent interactions may be a resource to compensate the scrambling produced by the ``bare'' Hamiltonian, thus ensuring a more dissipative dynamics and a larger burst.

An interesting avenue for future research is to investigate the quantum state of photons produced via many-body decay. The mirror configuration produces multi-photon states with similar metrological properties to Fock states~\cite{Gonzalez15,Paulisch19,Groiseau21,Perarnau_Llobet20,Molmer21}. However, in this configuration, photons need to be recombined into a single pulse, as they are emitted in both directions. This issue should be partially overcome at different lattice constants or in chiral waveguides. However, in these cases, dynamics may populate dark states, which are prevalent at low excitation densities ~\cite{Solano17,Albrecht_2019,Ferioli21_2,Fayard21}, trapping the last few photons in the pulse. Moreover, the direction of emission is initially random. However, stimulated emission may overcome this problem~\cite{Asaoka22,Kersten23}. Other promising lines of inquiry include the possibility of using measurement and feedback control on the output light to access entangled dark states and the investigation of non-Markovian effects in many-body decay \cite{Guo17,Sinha20,Carmele20,Arranz21}.

\begin{acknowledgments}
\textbf{Acknowledgments} -- We are thankful to J. T. Lee, R. Guti\'{e}rrez-J\'{a}uregui, W. Pfaff, W.-K. Mok, and L.-C. Kwek for stimulating discussions. We gratefully acknowledge support from the Air Force Office of Scientific Research through their Young Investigator Prize (grant No.~21RT0751), the National Science Foundation through their CAREER Award (No. 2047380), the A. P. Sloan foundation, and the David and Lucile Packard foundation. S. C.-L. acknowledges additional support from the Chien-Shiung Wu Family Foundation. A.A.-G. also acknowledges the Flatiron Institute, where some of this work was performed. We acknowledge computing resources from Columbia University's Shared Research Computing Facility project, which is supported by NIH Research Facility Improvement Grant 1G20RR030893-01, and associated funds from the New York State Empire State Development, Division of Science Technology and Innovation (NYSTAR) Contract C090171, both awarded April 15, 2010.
\end{acknowledgments}

\nocite{Malz22}
\nocite{Wiegner15}
\nocite{Wang22}
\nocite{Joshi22}

\bibliography{refs}

\onecolumngrid
\section{SUPPLEMENTAL MATERIAL}

\subsection{1. Calculation of the conditional statistics}
A necessary condition for observing a burst in a detector placed at the end of the waveguide is that the emission of photons into the waveguide increases at $t=0$. This condition can be written as $\tilde{g}^{(2)}(0)>1$, where the second-order correlation function reads
\begin{align}
    \tilde{g}^{(2)}(0) &= \frac{\sum\limits_{\nu=+,-} \sum\limits_{\mu=+,-,i} \Gamma_\nu \Gamma_\mu \braket{\jop^\dagger_\mu \jop^\dagger_\nu \jop_\nu \jop_\mu}}{\left(\sum\limits_{\nu=+,-}\Gamma_\nu \braket{\jop^\dagger_\nu\jop_\nu}\right) \left(\sum\limits_{\mu=+,-,i} \Gamma_\mu \braket{\jop^\dagger_\mu\jop_\mu}\right)} =1 + \frac{\sum\limits_{\nu=+,-}\Gamma_\nu^2 - N\ga \left( 2\ga + \gap\right)}{N^2\ga\left(\ga + \gap\right)}.
\end{align}

The condition $\tilde{g}^{(2)}(0)>1$ implies $\text{Tr}\left[\mathbb{\Gamma}^2\right]> N\ga (2\ga+\gap)$. The variance of $\{\Gamma_\nu\}$,

\begin{equation}
    \mathrm{Var}\left(\frac{\{\Gamma_\nu\}}{\ga}\right)=\frac{1}{N\ga^2}\sum_{\nu=1}^N(\Gamma_\nu^2-\ga^2),
\end{equation}

can be used to rewrite this expression as Eq. (7) of the main text.

\subsubsection*{1.1. Minimal burst condition for arbitrary spatial configurations}

Even when the analytical expression of the eigenvalues is not known, one can obtain a condition that guarantees a burst by finding a lower bound for $\sum_\nu\Gamma_\nu^2\equiv \text{Tr}\left[\mathbb{\Gamma}^2\right]$. The trace of $\mathbb{\Gamma}^2$ for an arbitrary spatial configuration is readily calculated as 
\begin{equation}
\begin{aligned}
\text{Tr}[\mathbb{\Gamma}^2]&=\sum_{i,j=1}^N \Gamma_{ij}^g \Gamma_{ji}^g =N^2 (\Gamma_R^2+\Gamma_L^2) +2\Gamma_R \Gamma_L \sum_{i,j=1}^N \cos (2\kk(z_i-z_j))\\
&=N^2 (\Gamma_R^2+\Gamma_L^2) +2\Gamma_R \Gamma_L \left[\left(\sum_{i=1}^N \cos (2\kk z_i)\right)^2+\left(\sum_{i=1}^N \sin (2\kk z_i)\right)^2\right]\geq  N^2 (\Gamma_R^2+\Gamma_L^2).
\end{aligned}
\end{equation}

Thus, choosing a set of parameters that satisfies $N (\Gamma_R^2+\Gamma_L^2)>\ga (2\ga+\gap)$, i.e., Eq. (10) in the main text, guarantees superradiance regardless of the arrangement of the emitters. If the emitters form an ordered array with lattice constant $\kk d= n \pi/N$ [i.e. at the degeneracy points in Fig.~2(a)], the bound is saturated. This can be intuitively understood from the fact that $\text{Tr}[\mathbb{\Gamma}^2]$ is related to the variance of $\{\Gamma_\nu\}$ which is minimized at degeneracy.

\subsubsection*{1.2. Enhancement of the emission of the third photon}

To confirm that $\tilde{g}^{(2)}(0)>1$ is a sufficient condition to predict a burst, we compute the conditional third order correlation function, which reads
\begin{align}
    \tilde{g}^{(3)}(0) &= \frac{\sum\limits_{\nu=+,-} \sum\limits_{\mu,\chi=+,-,i} \Gamma_\nu \Gamma_\mu \Gamma_\chi \braket{\jop^\dagger_\chi \jop^\dagger_\mu \jop^\dagger_\nu \jop_\nu \jop_\mu \jop_\chi}}{\left(\sum\limits_{\nu=+,-}\Gamma_\nu \braket{\jop^\dagger_\nu\jop_\nu}\right) \left(\sum\limits_{\mu=+,-,i} \Gamma_\mu \braket{\jop^\dagger_\mu\jop_\mu}\right)^2} \\
    &= 1 - \frac{6\Gamma_\text{1D}^2 + 8 \Gamma' \Gamma_\text{1D} + 3 \Gamma'^2}{N(\Gamma_\text{1D} + \Gamma')^2} + \frac{12\Gamma_\text{1D}^2 + 8 \Gamma' \Gamma_\text{1D} + 2 \Gamma'^2}{N^2 (\Gamma_\text{1D} + \Gamma')^2} \notag\\ & \qquad + \frac{\left(3N \Gamma_\text{1D} - 12 \Gamma_\text{1D} + 2(N-2) \Gamma'\right) \sum\limits_{\nu=+,-} \Gamma_\nu^2 + 2\sum\limits_{\nu=+,-} \Gamma_\nu^3}{N^3 \Gamma_\text{1D}(\Gamma_\text{1D} + \Gamma')^2}.
    \label{g3}
\end{align}	

Below we prove by contradiction that there is no combination of parameters such that $\tilde{g}^{(2)}(0)\leq 1$ while $\tilde{g}^{(3)}(0)>1$, so the emission of the third photon is never enhanced if the second was not. The inequality $\tilde{g}^{(2)}(0)\leq 1$ implies
\begin{equation}
\text{Tr}\left[\mathbb{\Gamma}^2\right]<N\ga (2\ga+\gap).
 \label{ineq1}
\end{equation}
Combining the above equation with $\tilde{g}^{(3)}(0)>1$ yields
\begin{equation}
2\text{Tr}\left[\mathbb{\Gamma}^3\right]>N^2\ga\gap (\ga+\gap)+2 N\ga (6\ga^2+6\ga\gap+\gap^2).
 \label{ineq2}
\end{equation}

After some algebraic manipulations, one finds
\begin{subequations}
 \begin{align}
\text{Tr}[\mathbb{\Gamma}^2] &= \sum_{i,j=1}^N \Gamma_{ij} \Gamma_{ji} = N^2(\Gamma_L^2 + \Gamma_R^2) + 2\Gamma_L\Gamma_R\sum_{i,j=1}^N \cos\left[2\kk (z_i-z_j)\right],\\
\text{Tr}[\mathbb{\Gamma}^3] &= \sum_{i,j,k=1}^N \Gamma_{ij} \Gamma_{jk} \Gamma_{ki} = N^3(\Gamma_L^3 + \Gamma_R^3) + 3N\Gamma_\text{1D}\Gamma_L \Gamma_R \sum_{i,j=1}^N \cos\left[2\kk (z_i-z_j)\right].
\end{align}
\end{subequations}

Therefore, the condition for the third emission to be the first one that is enhanced reduces to
\begin{equation}
F(N)\equiv N^2 +B N+C <0,
\label{fn}
\end{equation}
with 
\begin{subequations}
\begin{align}
B&=\left(\frac{\gap}{\ga}\right)^2-\frac{2\gap}{\ga}-6,\\
C&=2\left(\frac{\gap}{\ga}\right)^2+12+12\frac{\gap}{\ga}.
\end{align}
\end{subequations}
While $C>0$ for all possible values of the ratio $\ga/\gap$, $B>0$ only if $\gap/\ga>1+\sqrt{7}$. Thus, if $\gap/\ga>1+\sqrt{7}$, the inequality in Eq.~\eqref{fn}  is never satisfied. If $\gap/\ga<1+\sqrt{7}$,  the discriminant of $F(N)$ is negative, so $F(N)$ does not have real roots and hence it is never negative. We then conclude that  no combination of parameters yield $\tilde{g}^{(2)}(0)\leq 1$ while $\tilde{g}^{(3)}(0)>1$, so  $\tilde{g}^{(2)}(0)> 1$ is a sufficient condition to predict a burst.

\subsubsection*{1.3. Enhancement of emission into all channels}

The condition to have emission enhancement into all possible channels is calculated by imposing [30]
\begin{equation}
\mathrm{Var}\left(\frac{\{\tilde{\Gamma}_\nu\}}{\Gamma_0}\right) > 1,
\label{condAll}
\end{equation}
with $\tilde{\Gamma}_1=\gp+\gap$,  $\tilde{\Gamma}_2=\gm+\gap$, $\tilde{\Gamma}_{\nu>2}=\gap$ and $\Gamma_0=\ga+\gap$. After simplifying one finds
\begin{equation}
 \frac{N^2\left(\Gamma_L-\Gamma_R\right)^2}{2}+2 \Gamma_L\Gamma_R\frac{\sin^2 N\kk d}{\sin^2 \kk d}+\frac{\ga^2}{2}(N^2-2N)> N(\ga+\gap)^2,
\label{condAll2}
\end{equation}
which agrees with  conditions found in other works [72]. However, note that Eq.~\eqref{condAll2} is a condition for overall superradiant emission, as opposed to Eq.~(7) of the main text.

\subsection{2. Emergent translational symmetry at the degeneracy points}
Consider a finite and ordered array of $N$ emitters whose interactions are described through the matrix of coefficients $\Gamma_{ij}$. If the interactions satisfy periodic or antiperiodic boundary conditions (i.e.  if $\Gamma_{i,N+1}=\pm \Gamma_{i,1}$), the system becomes translationally invariant since one can identify particle 1 with particle $N+1$. Periodic boundary conditions (PBC) emerge for bidirectional waveguides at lattice constants $\kk d=2n\pi/N$, with $n\in \mathbb{N}$. Antiperiodic boundary conditions (APBC) are achieved for $\kk d=(2n+1)\pi/N$. Therefore, at $\kk d=n\pi/N$, the eigenvectors of $\mathbb{\Gamma}$ must obey Bloch's theorem. The jump operators thus take the form of annihilation operators for spin waves with momentum $k$ in the first Brillouin zone, i.e.,
\begin{equation}
 \hat{S}_{k}=\frac{1}{\sqrt{N}}\sum_{j=1}^N e^{-\ii k d j}\hge^j.
\end{equation}
The only two jump operators with finite decay rates are those that match the wavevector of the guided mode, i.e., $k=\pm \kk$. These are the left and right jump operators defined in Eq.~(11) of the main text. Since the system is mirror-symmetric, the two jump operators must have identical decay rates. Thus, the two non-zero eigenvalues of the $\mathbb{\Gamma}$ matrix become degenerate at $\kk d=n\pi/N$.

We can also demonstrate that the two non-zero eigenvalues are degenerate at $\kk d=n\pi/N$ by employing the analytical form of the collective jump operators. For the case of a bidirectional waveguide they read
\begin{equation}
\hat{\mathcal{O}}_\pm=\sqrt{\frac{\ga}{\Gamma_\pm}}\sum_{i=1}^N \mathcal{F}_\pm \left[\kk d \bigg(\frac{N+1}{2}-i\bigg)\right]\hge^i,
\label{eq:jumps}
\end{equation}
with $\mathcal{F}_{\{+(-)\}}[\cdot]=\cos[\cdot] (\sin[\cdot])$. Notice that $\hat{\mathcal{O}}_\pm$ differ by a $\pi/2$ phase shift. Due to the emergent translational symmetry at $\kk d=n\pi/N$, $\hat{\mathcal{O}}_\pm$ must have the same decay rates at these points.

\subsection{3. Channel competition and burst condition}

The degree of competition between the channels $\pm$ is dictated by the collective decay rates, $\Gamma_\pm$. The situation of minimal competition is that of a single bright channel. There are two scenarios that realize this:

\begin{itemize}
    \item A perfectly-chiral waveguide, for which $\Gamma_+=N\ga$ and $\Gamma_-=0$. While there is only one jump operator for any lattice constant, this situation is not exactly that studied by Dicke, due to the non-zero coherent evolution.
    \item  Mirror configuration for waveguides of arbitrary chirality. For bidirectional waveguides, the coherent interactions mediated by the Hamiltonian vanish, and the system reduces to the Dicke model. 
\end{itemize}

The burst condition in Eq.~(9) reduces to $(N-2)\ga>\gap$ for any of these two situations. Notice that the condition for an array with any lattice constant and $N\gg1$ coupled to a bidirectional waveguide reduces to $(N-4)\ga>2\gap$. In the limit of strong parasitic decay, constructive interference for $\kk d=n\pi$ makes a superradiant burst possible with fewer emitters than is needed for most other spatial configurations.

The scenario of maximal competition between decay channels occurs whenever their rates are identical. This happens only for bidirectional waveguides at the degeneracy points discussed in the last section. As seen in Fig. 2 of the main text, any level of chirality breaks the translation symmetry and opens a gap in the decay spectrum. In chiral waveguides and away from the ``mirror configuration'', the decay rates barely oscillate with $\kk d$ (as the sinusoidal term does not scale with $N$ and can be ignored) and saturate to $\gp\simeq N\text{max}(\gl,\gr)$ and  $\gm\simeq N\text{min}(\gl,\gr)$.

\subsection{4. Chirality induced by jumps}

The conditioned correlation function [Eq.~(6) of the main text] can be expressed in terms of the directional correlation functions [Eqs.~(12a) and~(12b) of the main text] as

\begin{equation}
\tilde{g}^{(2)}(0)=\frac{2 \gl\gr}{\ga^2}\tilde{g}_{LR}^{(2)}(0)+\frac{\gr^2+\gl^2}{\ga^2}\tilde{g}_{LL}^{(2)}(0).\label{g2_dir}
\end{equation}

To prove this, we have used  $\tilde{g}_{LL}^{(2)}(0)=\tilde{g}_{RR}^{(2)}(0)$ and $\tilde{g}_{LR}^{(2)}(0)=\tilde{g}_{RL}^{(2)}(0)$.

Directional correlation functions  can be defined for more than two jumps and should show emergent chirality beyond the second photon. Below, we ignore Hamiltonian evolution, since we want to emphasize the effect of jumps on directional emission. The probability that the $n$th photon is emitted in the $\alpha$ direction ($\alpha=L,R$) if the previous $n-1$ photons were emitted to the left is proportional to
\begin{equation}
     \tilde{g}_{L\alpha}^{(n)}(0)=\frac{\braket{{\jop_L}^{\dagger n-1}\jop^\dagger_\alpha\jop_\alpha{\jop_L}^{n-1}}}{\braket{\jop^\dagger_L\jop_L}^{n-1}\braket{\jop^\dagger_\alpha\jop_\alpha}}.
\end{equation}

Using Eq.~(11) of the main text, and simplifying, we find [73]
\begin{subequations}
\begin{eqnarray}
     \tilde{g}_{LL}^{(n)}(0)&&=\frac{1}{N^n}\sum_{\substack{\sigma_1,... ,\sigma_n=1\\ \sigma_1<\sigma_2<...<\sigma_n}}^N\left|n! \quad  e^{\ii\kk d\sum\limits_{m=1}^n \sigma_m}\right|^2=\frac{n!N!}{N^n(N-n)!},\label{gll_njump}\\
     \tilde{g}_{LR}^{(n)}(0)&&=\frac{(n-1)!^2}{N^n}\sum_{\substack{\sigma_1,... ,\sigma_n=1\\ \sigma_1<\sigma_2<...<\sigma_n}}^N\left|\sum_{m=1}^n e^{-2\ii \kk d\sigma_m}\right|^2=\frac{(n-1)! N!}{(N-n)! N^n (N-1)}\left(N-n+\frac{n-1}{N}\frac{\sin^2 N\kk d}{\sin^2 \kk d}\right).
     \label{glr_njump}
\end{eqnarray}
\end{subequations}
Here, the sum over $\sigma_1$, $\sigma_2$, ... $\sigma_n$  accounts for all the possible final states after the $n$ detection events, i.e., all possible choices of $n$ emitters to be the ones that emitted a photon and decayed to the ground state. For each final state, we need to coherently sum the contribution of all the quantum paths that lead to it [33, 39]. In the case of  $\tilde{g}_{LL}^{(n)}(0)$, the maximum constructive interference between paths is achieved since all $n!$ paths have the same phase. On the other hand, the quantum paths for each final state in $\tilde{g}_{LR}^{(n)}(0)$ do not have the same phase and hence do not achieve maximum constructive interference.

The ratio $r(n)=\tilde{g}_{LL}^{(n)}(0)/\tilde{g}_{LR}^{(n)}(0)$ is a measure of the emergent chirality after the $n$th detection event in the absence of Hamiltonian dynamics. From Eqs.~\eqref{gll_njump} and~\eqref{glr_njump},

 \begin{equation}
     r(n)=\frac{nN(N-1)\sin^2\kk d}{(N-n)N\sin^2\kk d+(n-1)\sin^2 N\kk d}.\label{rn}
 \end{equation}
 
\begin{figure*}
\centering
\includegraphics[width=0.5\textwidth]{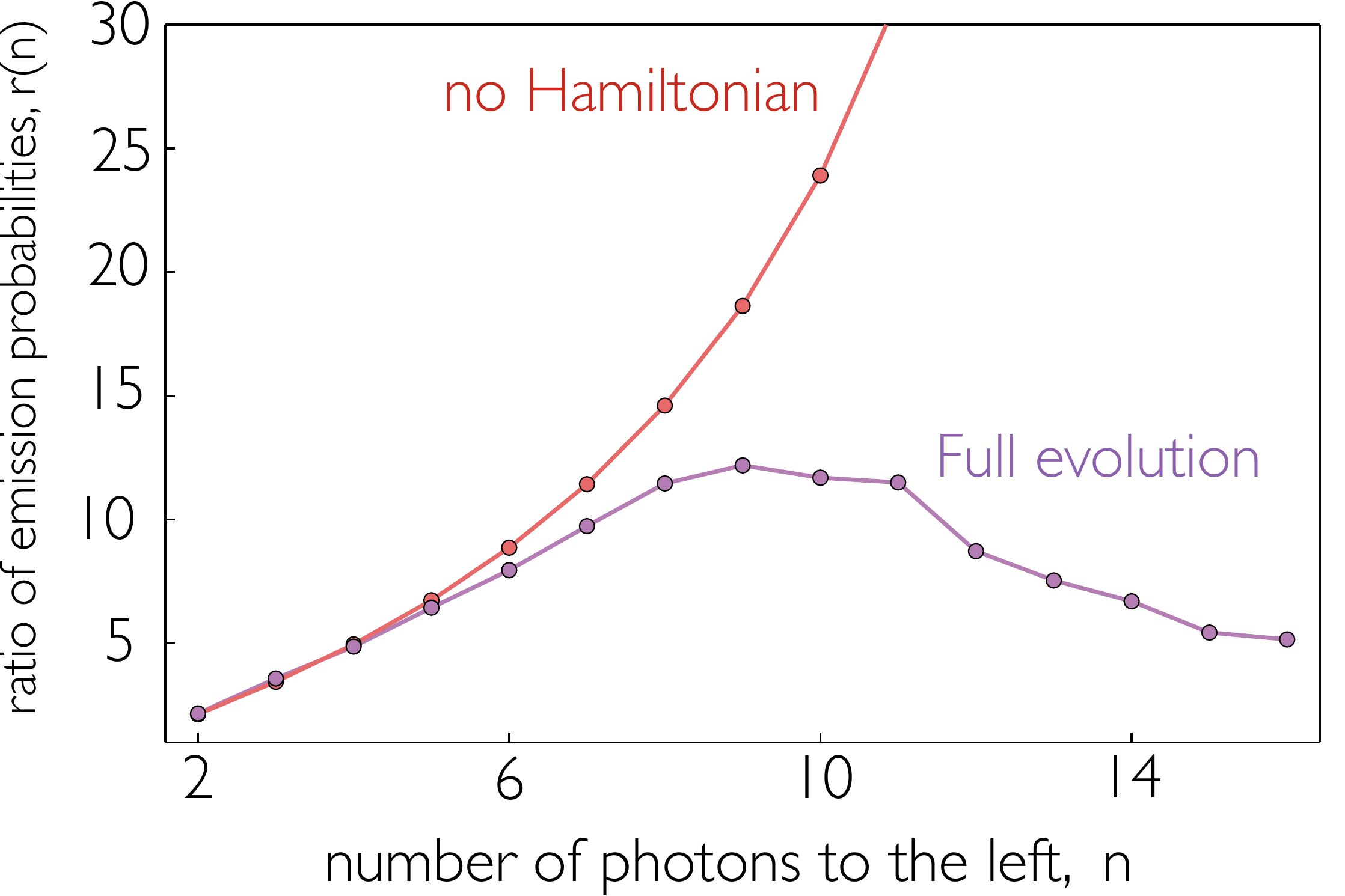}
\caption{Ratio of probabilities of detecting a photon to the left or right after $n-1$ photons have been emitted to the left, versus photon number. The full evolution is shown in purple, and the red line depicts the analytical ratio obtained by only considering jumps.}
\label{fig:expratio}
\end{figure*}

Fig.~\ref{fig:expratio} shows $r(n)$ with and without Hamiltonian evolution. For the former, the probabilities are calculated numerically using quantum trajectories. As noted in the main text, Hamiltonian evolution scrambles the emitters' phases and suppresses chirality enhancement.

For many atoms $N\gg 1$ that are not in the mirror configuration (thus $\sin^2\kk d\simeq \sin^2 N\kk d$), the ratio reduces to $r(n)\sim n$ for $n\ll N$, both for the full and the approximated dynamics. This means that if $n$ photons have been emitted into one direction, the next photon is $n$ times more likely to be emitted in that same direction.

\subsection{5. Imbalance statistics with parasitic decay}
\begin{figure*}[h]
\centering
\includegraphics[width=\textwidth]{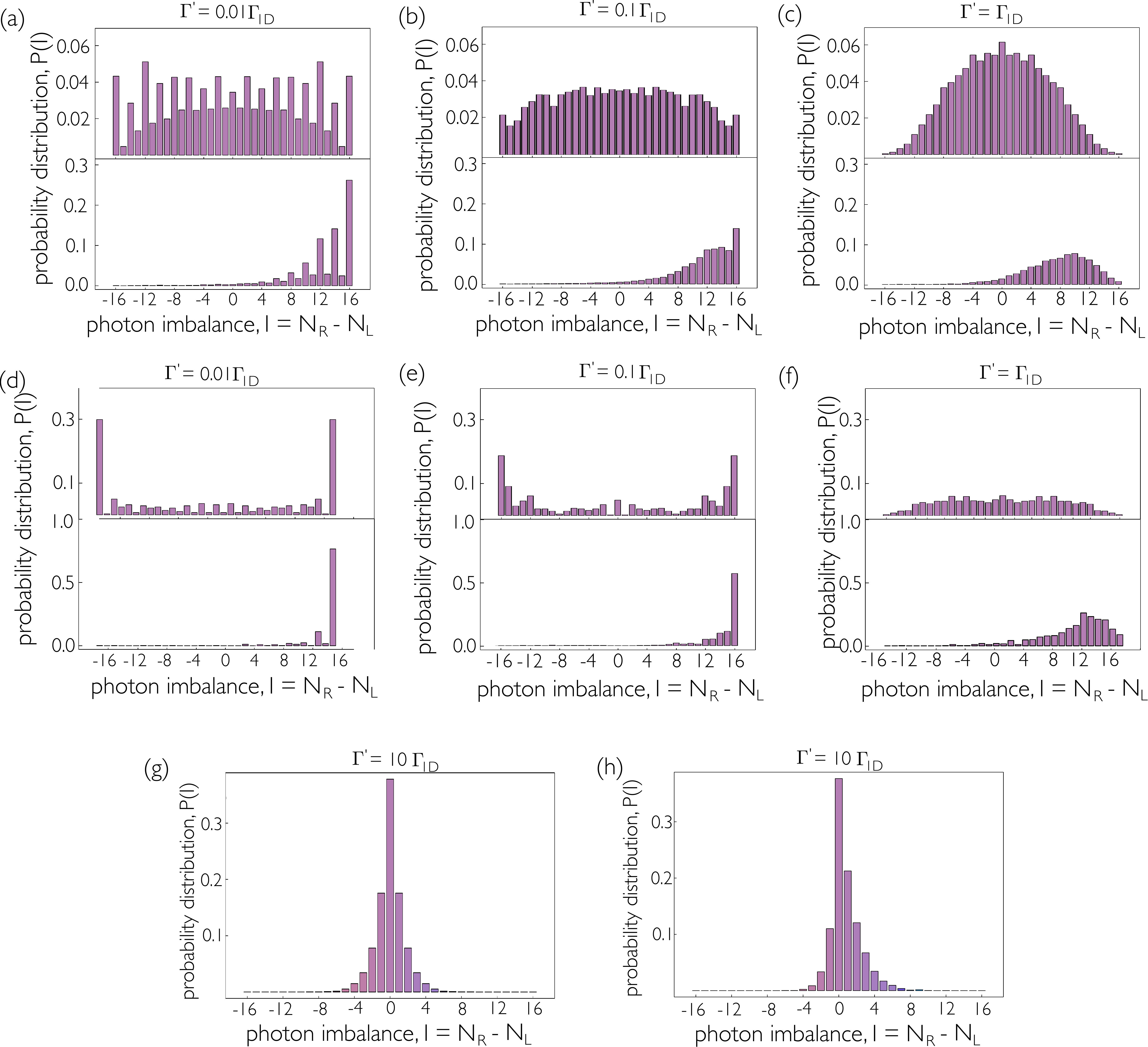}
\caption{Imbalance statistics on the emission direction for 16 emitters with $\gap=0.01\ga$, $\gap=0.1\ga$ and $\gap=\ga$ for $\ga t=100$ (a-c) and  $\ga t=1$ (d-f). For each subplot, the top and bottom panels show respectively the probability distribution for a bidirectional and a chiral waveguide ($\gr=3\gl$). (g-h) Show the imbalance statistics with $\gap=10\ga$ for a bidirectional and a chiral waveguide for $\ga t=1$. }
\label{fig:imbalances_robustness}
\end{figure*}

Here we investigate in more detail the effects of parasitic decay in the photon imbalance probability distributions shown in Fig.~3. Figure~\ref{fig:imbalances_robustness} (a)-(f) shows the probability distributions for three different ratios of $\gap/\ga$ for late and early times. $I\in\{-N, -N+2, ..., N\}$ when the parasitic decay is zero, as the number of emitted photons into the left and right-propagating modes must satisfy $N_R+N_L=N$. If $\gap\neq 0$, $I$ can take any value as long as $N_R+N_L\leq N$. The enhancement in the probability of maximum imbalance survives parasitic decays achievable in state-of-the-art experimental setups with superconducting qubits, where $\Gamma'\simeq 0.01\Gamma_\text{1D}$~[15], and quantum dots, where $\Gamma'\simeq 0.1\Gamma_\text{1D}$~[13]. If $\ga\ll \gap$, correlations are washed out, and there is a very suppressed probability of large imbalances, as shown in Fig.~\ref{fig:imbalances_robustness} (g) and (h). 

Correlated decay with $\gap=0$ produces a nearly flat $\mathcal{P}(I)$ for large times. By approximating $\mathcal{P}(I) \sim 1/(N+1)$, we note that the variance scales quadratically with the number of emitters
\begin{equation}
    \text{Var}(I)_\text{flat}=\frac{N(N+2)}{3}.
\end{equation}
In contrast, if there is no correlation between the direction of emission of each photon and $\gap\neq 0$, the decay can be described in terms of a multinomial probability distribution. Each photon has a probability $p_{L/R}=\Gamma_{L/R}/(\ga+\gap)$ of being emitted to the left/right, and a probability $p_{ng}=\gap/(\ga+\gap)$ of being emitted to non-guided modes. The variance in this case scales linearly with the number of emitters,
\begin{equation}
    \text{Var}(I)_\text{indep}=N\frac{\ga}{\ga+\gap}.
\end{equation}

\begin{figure*}[h]
\centering
\includegraphics[width=1\textwidth]{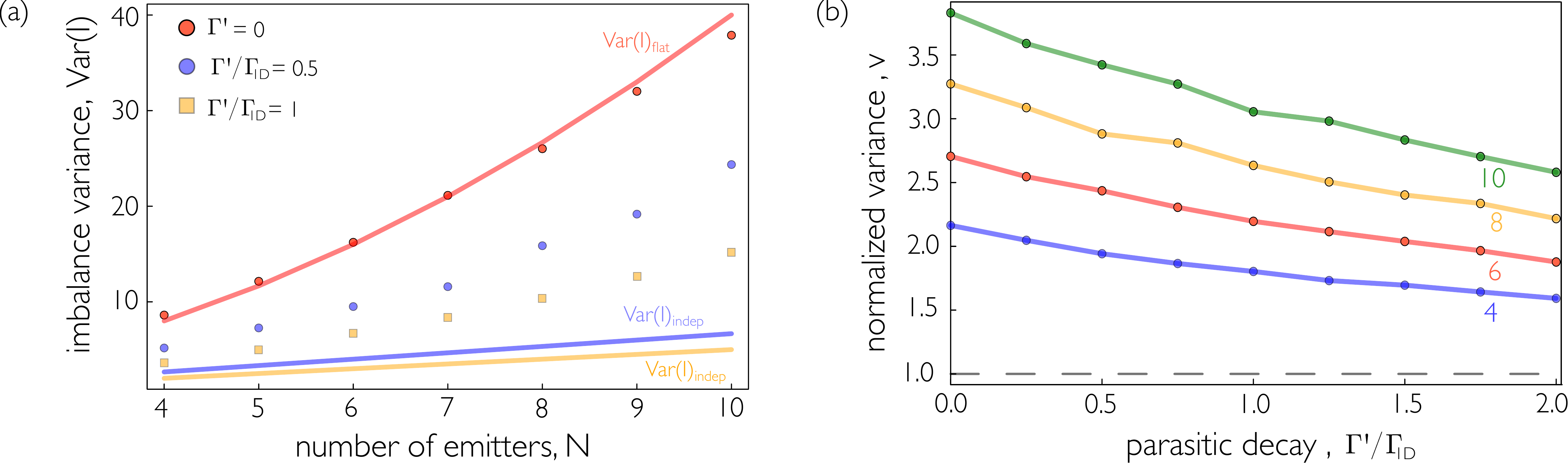}
\caption{(a) Scaling of the variance of the imbalance distribution  vs the number of emitters. The dots show numerical results for 4, 6, 8, and 10 emitters, while the lines show the variance for the flat and the corresponding multinomial distributions for the cases $\gap\neq 0$. (b) Ratio $v=\text{Var}(I)/\text{Var}(I)_\text{indep}$ for different number of emitters.}
\label{fig:scaling}
\end{figure*}

Fig.~\ref{fig:scaling} shows that for low parasitic decays, the scaling of the variance obtained from the decay is quadratic. As $\gap$ increases, correlations between the emission direction of different photons is partially destroyed. For the large $\gap$ the scaling approaches linear, although the variances are still larger than those predicted from purely uncorrelated emission. The degree of correlation in emission direction is captured by the ratio $v=\text{Var}(I)/\text{Var}(I)_\text{indep}$. If $v\gg1$ then there are strong correlations, while $v\sim 1$ signifies nearly uncorrelated emission. Even for $\Gamma'\geq\Gamma_\text{1D}$, the direction of emitted photons shows correlated behavior, as $v > 1$ and grows with $N$.

\subsection{6. Many-body superradiance in giant atoms}
\begin{figure*}
\centering
\includegraphics[width=0.9\textwidth]{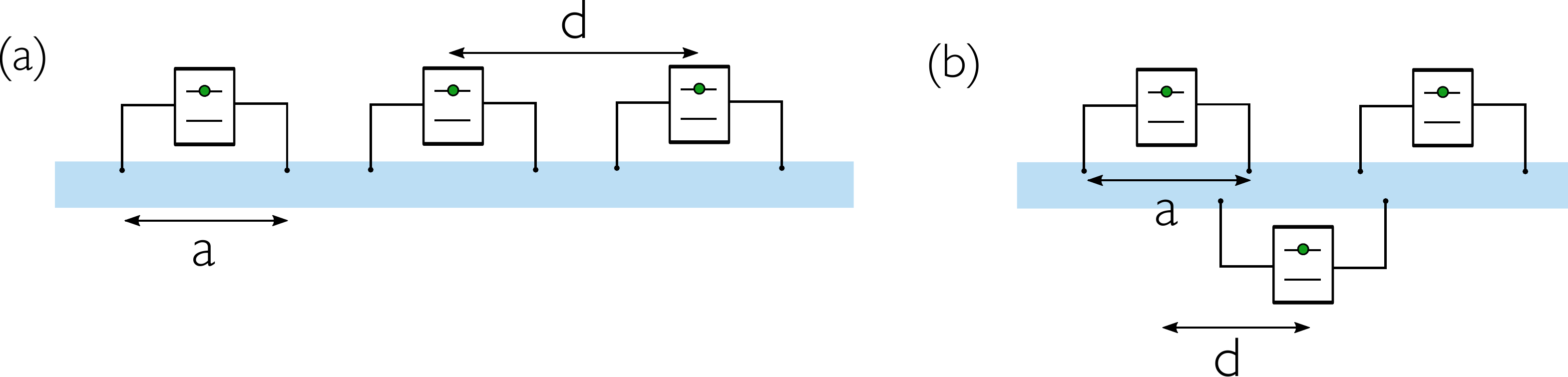}
\caption{$N$ giant atoms with two connection points in the (a) separated and (b) braided configuration.}
\label{fig:sketch}
\end{figure*}

Here we consider an array of $N$ giant atoms with two connection points in the separated and braided configurations (see Fig.~\ref{fig:sketch}). The distance between the coupling points of the same atom and the distance between consecutive giant atoms are denoted as $a$ and $d$, respectively. In the separated configuration, $d>a$, while in the braided configuration considered here, $2d>a>d$.  For simplicity, we assume a bidirectional waveguide and that the coupling strength  $\gamma_\text{1D}$ is the same for all connection points. We also assume that all the atoms have a parasitic decay to non-guided modes at a rate $\Gamma'$.

The atomic dynamics is described by the master equation~[55]
\begin{equation}
\dot{\ra}=-\frac{\ii}{\hbar}\left[H,\ra\right]+\mathcal{L}_g[\ra]+\mathcal{L}_{ng}[\ra],
\label{eq:mastergiant}
\end{equation}

with $\mathcal{L}_\alpha[\ra]=\sum\limits_{i,j=1}^N\frac{\Gamma^\alpha_{ij}}{2}\left(2\hge^j\ra\heg^i-\ra\heg^i\hge^j-\heg^i\hge^j\ra\right)$ as in the main text, and
\begin{subequations}
\begin{align}
    H&=\sum_{i,j=1}^{N} J_{ij}\heg^i\hge^j,\\
    J_{ij}&=\sum_{n,m=1}^2\frac{\gamma_\text{1D}}{2}\sin \phi_{(in,jm)},\label{JM}\\
    \Gamma^g_{ij}&=\sum_{n,m=1}^2 \gamma_\text{1D} \cos \phi_{(in,jm)}. \label{GammaM}
\end{align}
\end{subequations}
Here, $\phi_{(in,jm)}\equiv |\kk (z_{i,n}-z_{j,m})|$ is the phase acquired by light when traveling from the connection point $n$ of atom $i$ to the connection point $j$ of atom $m$. Connection points belonging to the same atom can interfere and enhance or suppress the individual decay~[55]. From  Eq. \eqref{GammaM}, the individual decay to the waveguide is $\Gamma_\text{1D}(a)=2\gamma_\text{1D} (1+\cos \kk a)$. By changing $a$, the individual decay can be tuned from $\Gamma_\text{1D}(a)=0$ to $\Gamma_\text{1D}(a)=4\gamma_\text{1D}$. 

We investigate the decay for a fully-inverted array of giant atoms, and compare it to the case of regular emitters with a single connection point. In accordance with the main text, the normalized emission rate into the waveguide for giant atoms is defined as
\begin{equation}
    R(t)=\frac{1}{N \Gamma_\text{1D}(a)}\sum_{\nu }\Gamma_\nu \braket{\hat{\mathcal{O}}_\nu^\dagger\hat{\mathcal{O}}_\nu},
\end{equation}
where $\{\mathcal{O}_\nu\}$ and $\{\Gamma_\nu \}$ are the collective jump operators and decays obtained from diagonalizing the $\mathbb{\Gamma}$ matrix obtained from Eq.~\eqref{GammaM}.

\begin{figure*}
\centering
\includegraphics[width=0.6\textwidth]{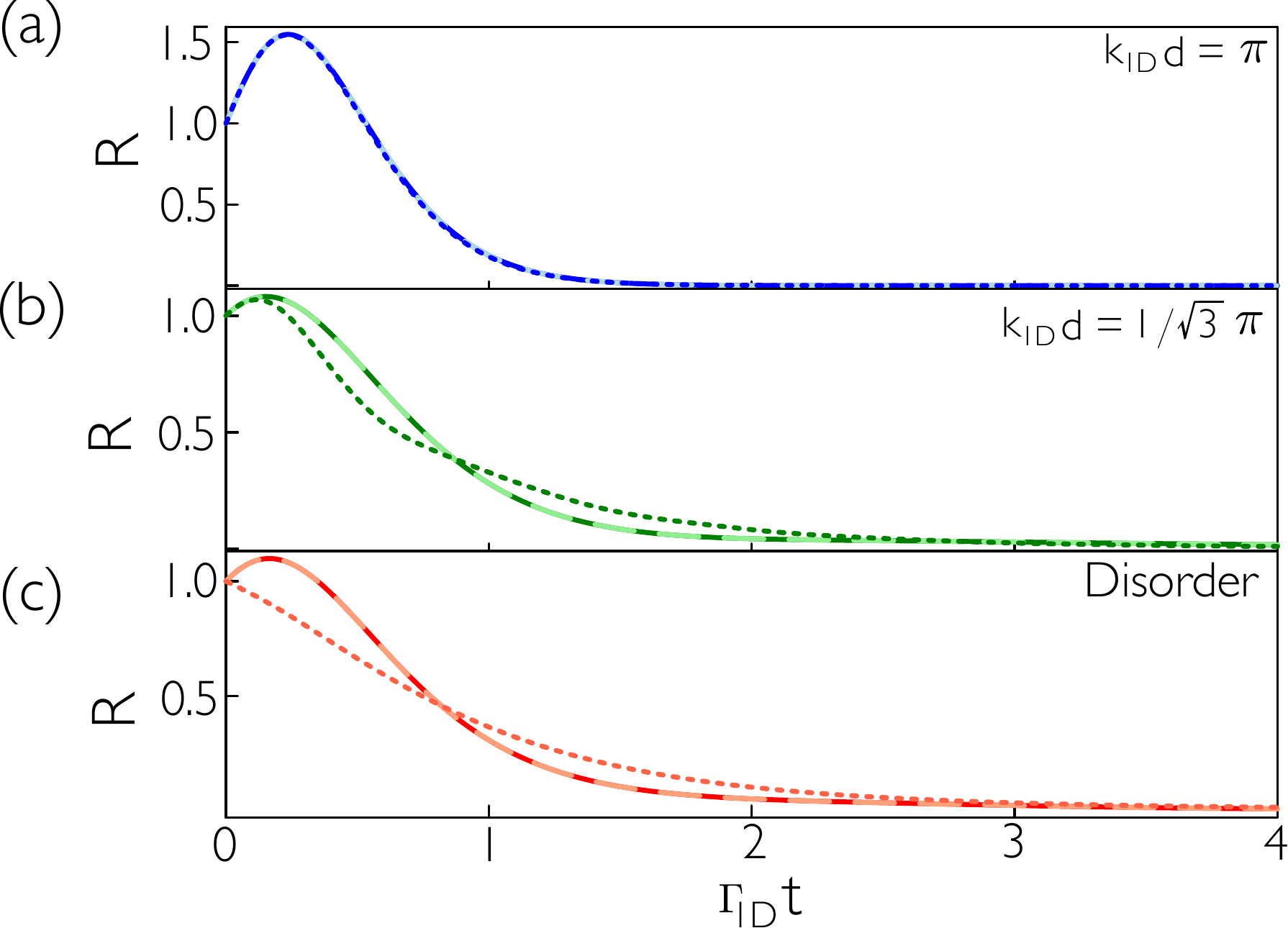}
\caption{Burst for giant atoms in the separated (dashed light lines) and braided (dashed lines) configurations, and for regular emitters (solid dark lines) for three different choices of atom positions. (a,b) Burst for an ordered array with $\kk d=\pi$ and $\kk d=\pi/\sqrt{3}$. In each case, the distances between connection points are  $a= 1/3 d$ for separated atoms and $a=3/2 d$ for braided atoms. (c) Burst of a completely disordered ensemble with $\kk a=0.1$ for the separated configuration  and $\kk a=0.99\pi$ for the braided configuration. The distance between consecutive atoms $\{d_{i,i+1}\}$ is chosen at random, with the only constraint that $d_{i,i+1}>a$ for separated atoms and $2d_{i,i+1}>a>d_{i,i+1}$ for the braided configuration. For all plots, $\gap=0$.}
\label{fig:bursts}
\end{figure*}

Figure~\ref{fig:bursts} shows the normalized emission rate for six giant atoms in separated and braided configurations and for six regular emitters. 
The burst produced by the array of separated giant atoms matches the burst produced by an array of regular emitters if $\kk a$ and $\gamma_\text{1D}$ such that $\Gamma_\text{1D}(a)=\Gamma_\text{1D}$, while the braided configuration shows discrepancies, especially in the long-time behavior. To understand this, we  note that  the  coefficients in Eq.~\eqref{GammaM} can be rewritten as 
\begin{equation}
    \Gamma^g_{ij}=\Gamma_\text{1D}(a) \cos \kk d|i-j|
\end{equation}
for both configurations. For the separated configuration, the coherent interaction coefficients in Eq.~\eqref{JM} admit a similar form,
\begin{equation}J_{ij}^\text{(sep.)}=\begin{cases}\frac{\Gamma_\text{1D}(a)}{2} \sin \kk d|i-j|\text{ if } |i-j|>0\\
\gamma_\text{1D} \sin \kk a \text{ if } i=j
\end{cases}
\end{equation}
The dynamics for atoms in the separated configuration is hence identical to that of regular emitters with a modified decay rate $\Gamma_\text{1D}(a)$ and a global frequency shift. This simplification is possible only for beyond-nearest-neighbor interactions for atoms in the braided configuration of Fig.~\ref{fig:sketch}(b),
\begin{equation}
    J_{ij}^\text{(braid)}=\begin{cases}\frac{\Gamma_\text{1D}(a)}{2} \sin \kk d|i-j|\text{ if } |i-j|>1\\
   \gamma_\text{1D}(\sin \kk d+\sin \kk a \cos \kk d)\text{ if } |i-j|=1\\
    \gamma_\text{1D} \sin \kk a \text{ if } i=j
    \end{cases}
\end{equation}
The burst produced by an array of braided giants atoms thus deviates from the normal emitter case due to these modified coherent interactions. In particular, if  $\kk a= \pi$, the system becomes completely decoherence-free since   $\Gamma^g_{ij}=0$ $\forall$ $i,j$, while the nearest-neighbor coherent interactions can be nonzero~[55]. If $\kk a\sim \pi$, the collective decay rates can be small when compared to the nearest-neighbor coherent interactions. As seen in Fig.~\ref{fig:bursts}(c), this can dephase the system and even quench the burst.

The burst far from this decoherence-free zone can be predicted by using the condition in  Eq. (7) of the main text. As in the case of the normal emitters, there are only two bright channels for both of the configurations, with decay rates
\begin{equation}
\Gamma_{\pm}^\text{(giant)}=\frac{\Gamma_\text{1D}(a)}{2}\left(N\pm\frac{\sin N\kk d}{\sin \kk d}\right).
\end{equation}
This is Eq.~(8) from the main text for the case of a bidirectional waveguide with $\Gamma_\text{1D}\rightarrow \Gamma_\text{1D}(a)$. The minimal burst condition deduced from Eq.~(7) reads
\begin{equation}
    \frac{N}{2}+\frac{\sin N(\kk d)^2}{2 N\sin(\kk d)^2}>2+\frac{\Gamma'}{\Gamma_\text{1D}(a)}.
\end{equation}

\begin{figure*}
\centering
\includegraphics[width=0.6\textwidth]{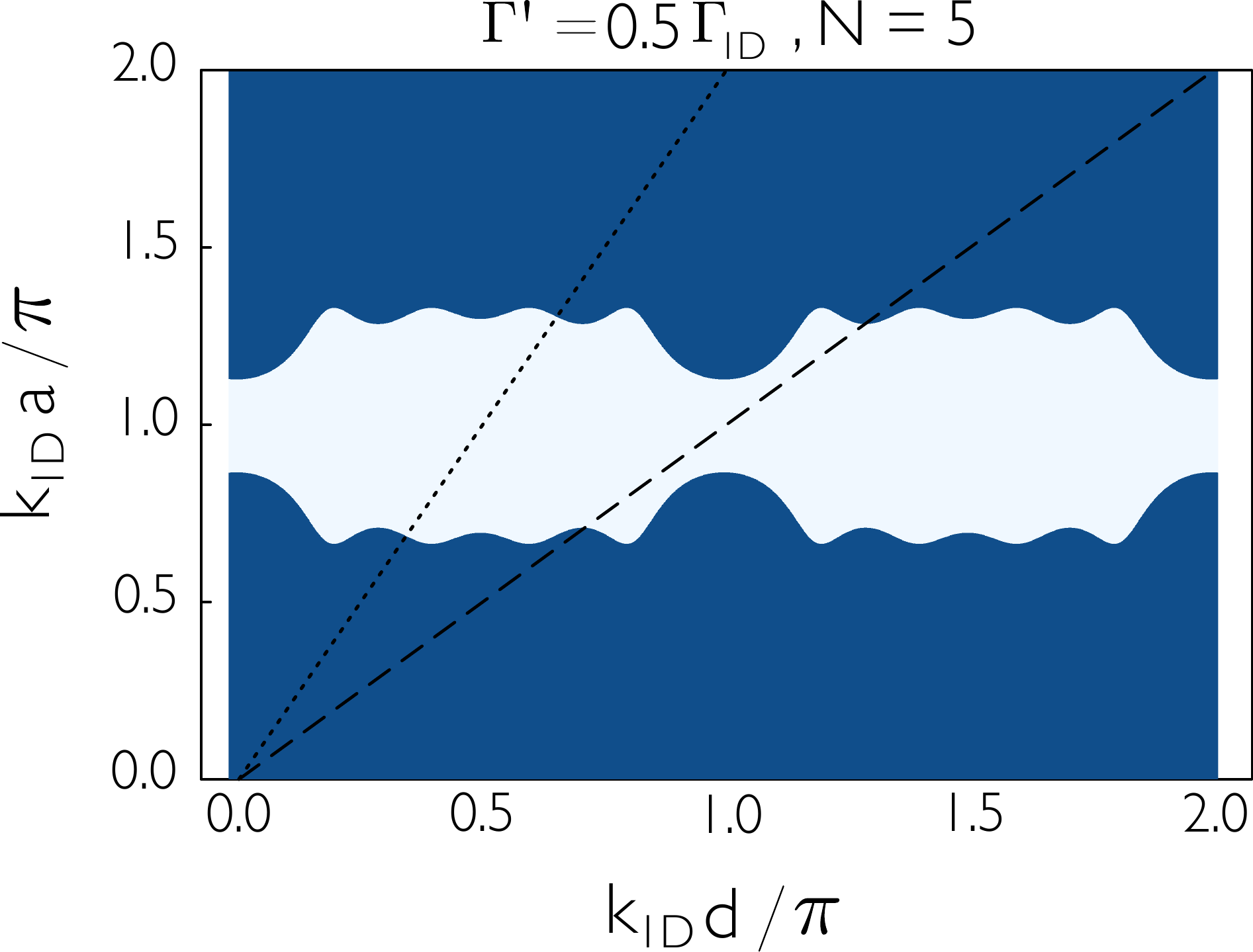}
\caption{Crossover between burst (dark blue) and no-burst (pale blue) regions for giant atoms in the separated (region below the $a=d$ line) and braided configuration (region between the lines $a=d$ and $a=2d$).}
\label{fig:cross}
\end{figure*}

Figure~\ref{fig:cross} shows a crossover diagram between burst and no-burst regions in the $a$-$d$ space for 5 giant atoms with a parasitic decay $\gap=0.1\ga$, which is in the range of experimentally-relevant parameters~[56]. The regions above and below the line  $a=d$ belong to braided and separate configurations, respectively. The region above the line $a=2d$ corresponds to configurations where atoms braid with beyond-nearest neighbors.

Braiding atoms enables the control of the ratio between coherent interactions and collective decay rates.  This approach could potentially be used to minimize coherent interactions, thus minimizing the scrambling of phases in between jumps and enhancing the probability of emitting all photons in the same direction. Other intriguing questions to explore are the interplay between collective decay and non-Markovian effects~[57-59] and chiral emission~[74, 75].

\subsection{7. Details on numerical evolution}

Here we give more details on the numerical evolution and the choice of parameters. The number of trajectories used for Figs.~1 and ~3 (on the main text) , and SM Fig. 2 is detailed in Table~\ref{table1}. 

\begin{table}[H] 
 \caption{\label{table1} Number of trajectories used for Figs. ~1 and ~3 of the main text, and SM Fig. 2}
 \begin{ruledtabular}
\begin{tabular}{lll}
                        & Bidirectional & Chiral \\
 Fig. 1               & 2,000         & 2,000       \\                       
Fig. 3(b)               & 12,000         & 12,000       \\
Fig. 3(c) and Fig. 3(d) & 20,000        & 13,000     \\
SM Fig. 2                &  5,000       &   8,000  

 \end{tabular}
 \end{ruledtabular}
\end{table}

The calculations for Figs. 2(b) and 2(c) are performed in the constrained Hilbert space composed of states with at most two atoms in $\ket{g}$. A burst is predicted if the emission rate at the final time $N\ga t_\text{fin}=10^{-5}$ is larger than at $t=0$.

The probability in Fig. 2(c) is computed with 100 random configurations.  The numerical boundary for each $N$ is defined as the minimum $\ga/\gap$ for which a burst occurs in every configuration. If $k_\text{1D}z_\text{max}\ll 2\pi$, all the emitters are approximately coupling to the waveguide at the same point, and the system is effectively in the mirror configuration. To make sure we are just considering truly disordered configurations the emitter positions are chosen at random from $[0,z_\text{max}]$ with $k_\text{1D}z_\text{max}\gg2\pi$.
In Figs.~3(c) and (d) $\pi$ over an irrational number was chosen to avoid arrays that are commensurable with the photon wavelength $\lambda_\text{1D}$, like at the degeneracy points, and show the behavior of a generic array.

\end{document}